\documentclass[sigconf]{acmart}
\PassOptionsToPackage{hyphens}{url}
\usepackage[hyphens]{url}  
\usepackage{graphicx} 
\usepackage{graphicx}  
\usepackage{array, boldline, makecell, booktabs}
\usepackage{colortbl}
\definecolor{lgrey}{RGB}{140,140,140}
\definecolor{mred}{RGB}{0, 128, 255}
\usepackage{pbox}
\usepackage{multirow}  
\newcolumntype{P}[1]{>{\centering\arraybackslash}p{#1}}
\usepackage{enumerate}
\usepackage[shortlabels]{enumitem}
\usepackage{hyperref}
\usepackage{url}

\AtBeginDocument{%
  \providecommand\BibTeX{{%
    \normalfont B\kern-0.5em{\scshape i\kern-0.25em b}\kern-0.8em\TeX}}}

\copyrightyear{2021} 
\acmYear{2021} 
\setcopyright{acmcopyright}\acmConference[CHI '21]{CHI Conference on Human Factors in Computing Systems}{May 8--13, 2021}{Yokohama, Japan}
\acmBooktitle{CHI Conference on Human Factors in Computing Systems (CHI '21), May 8--13, 2021, Yokohama, Japan}
\acmPrice{15.00}
\acmDOI{10.1145/3411764.3445449}
\acmISBN{978-1-4503-8096-6/21/05}

\begin{document}

\title{Exploring Semi-Supervised Learning \\ for Predicting Listener Backchannels}

\author{Vidit Jain}
\authornotemark[1]
\email{vidit17370@iiitd.ac.in}
\affiliation{
  \institution{IIIT-Delhi}
  \city{New Delhi, India}
}

\author{Maitree Leekha}
\email{ maitreeleekha\_bt2k16@dtu.ac.in}
\affiliation{
  \institution{Delhi Technological University}
    \city{New Delhi, India}}
\authornote{The authors contributed equally, and wish they be regarded as joint first authors.}

\author{Rajiv Ratn Shah}
\email{rajivratn@iiitd.ac.in}
\affiliation{
  \institution{IIIT-Delhi}
  \city{New Delhi, India}
}
\authornote{Jainendra Shukla and Rajiv Ratn Shah are partly supported by the Infosys Center for AI and Center for Design and New Media at IIIT Delhi.}
\authornote{Rajiv Ratn Shah is also partly supported by the ECRA Grant (ECR/2018/002776) by SERB, Government of India.}

\author{Jainendra Shukla}
\email{jainendra@iiitd.ac.in}
\affiliation{
  \institution{IIIT-Delhi}
  \city{New Delhi, India}
}
\authornotemark[2]

\begin{abstract}
Developing human-like conversational agents is a prime area in HCI research and subsumes many tasks. Predicting listener backchannels is one such actively-researched task. While many studies have used different approaches for backchannel prediction, they all have depended on manual annotations for a large dataset. This is a bottleneck impacting the scalability of development. To this end, we propose using semi-supervised techniques to automate the process of \textit{identifying} backchannels, thereby easing the annotation process. To analyze our identification module's feasibility, we compared the backchannel \textit{prediction} models trained on (a) manually-annotated and (b) semi-supervised labels. Quantitative analysis revealed that the proposed semi-supervised approach could attain 95\% of the former's performance. Our user-study findings revealed that almost 60\% of the participants found the backchannel responses predicted by the proposed model more natural. Finally, we also analyzed the impact of personality on the type of backchannel signals and validated our findings in the user-study.
\end{abstract}

\ccsdesc[300]{Human-centered computing~User studies}
\ccsdesc[500]{Human-centered computing~Empirical studies in HCI}
\ccsdesc[500]{Computing methodologies~Semi-supervised learning settings}

\keywords{Conversational Agents, Backchanneling, Multimodal analysis.}

\maketitle

\section{Introduction \& Relation to Prior work}
\label{intro_lr}
Human conversations, even the most casual ones, have a lot of complexity associated with them. Two people actively engaged in a conversation frequently respond to each other, not only with respect to the content of the conversation, but also to behavioral aspects, such as facial expressions and prosody. Developing Embodied Conversational Agents (ECAs) \cite{cassell2000embodied} and spoken dialogue systems capable of incorporating these complex elements to converse naturally is a challenging task and has been a constant focus of the Artificial Intelligence and Human Computer Interaction research communities. Of these complex conversational constructs, dyadic components like listener backchannels are among the most crucial for modeling virtual humans and are also the main focus of this study\footnote{A teaser video demonstrating our work can be found in the video figure section of the submission. It illustrates a virtual listener that emits backchannels to the speaker's context using the models proposed with this work.}.  

In a peer-to-peer conversation, a backchannel occurs when one of the participants is speaking, and the other (the listener) interjects a short response to the former \cite{white1989backchannels}. These responses do not interrupt the flow of the conversation; rather, they convey the listener's state-of-mind about the speaker's dialog. They also reflect cooperation and understanding between the two parties \cite{heinz2003backchannel}. Backchannels can be verbal, non-verbal (visual) or both. Vocalisations like 'hmm' or 'uh-huh', gestures such as head nods or head shakes, and a combination of verbal and non-verbal responses are common examples of backchannels.

In the past few years, the research community has shown a keen interest in modeling the listener's backchanneling behavior. A large number of such studies on backchannel prediction have focused on the use of rule-based classifiers. Ward \cite{ward1996}, Truong \textit{et al.} \cite{truong2010rule}, Ward and Tsukahara \cite{ward2000prosodic} utilized different acoustic features of the speaker such as pitch, pausal information, etc. to predict backchannel opportunities. A recent study by Park \textit{et al.} \cite{park2017} on backchannel prediction for children also used similar prosodic features and hand-crafted rules. Expanding the feature set, Moubayed \textit{et al.} \cite{Moubayed2009} used both visual and prosodic speaker features.

Transitioning towards data-driven automatic prediction from hand-crafted rules, Solorio \textit{et al.} \cite{Solorio2006ProsodicFG} used prosodic features with locally weighted linear regression to predict backchannel opportunities. Morency \textit{et al.} \cite{morency2010} used   multimodal (visual and acoustic) speaker features and a hidden markov model to predict backchannels. More recently, many researchers also used deep learning techniques for predicting backchannels. Ruede \textit{et al.} \cite{Ruede2019} used Long short-term memory (LSTM) based model with acoustic features. Hara \textit{et al.} \cite{Hara2018PredictionOT} also used LSTMs to predict turn-taking, filler words, and backchannels in a multitask learning paradigm. In a recent study, Goswami \textit{et al.} \cite{goswami2020social} used state of the art machine learning and deep learning-based time-series classification techniques to model backchannel opportunities in children with multimodal features. Inspired by prior work, we too explore the use of machine learning and deep learning based time series models and include several similar multimodal features in our analysis.


All these studies, although well-founded, have the following limitations:

\begin{enumerate}
    \item Most of them have focused only on the backchannel opportunity prediction task, with a very few involving the next step of the problem \textit{i.e.,} predicting the type of backchanneling signals. 
    
    \item All studies in the literature have relied on data with annotated backchannel instances during their modeling phase, where they develop a backchannel predictor using the speaker context. For instance, the datasets used by prior works, including the Switchboard Dialog Act Corpus (SwDA) \cite{jurafsky1997switchboard}, Iraqi Arabic \cite{ward2006case}, P2PStory \cite{singh2018p2pstory}, data collected by Morency \textit{et al.}  \cite{morency2010}, and many others, had to be manually annotated for listener backchannels by multiple coders. This annotation process can be extremely time-consuming, depending on the amount of data present. Furthermore,  this approach also does not scale well when trying to expand the scope of development by collecting more data. For instance, in many studies, including the ones cited above, the datasets used cover a relatively small population size. The size of the dataset used is primarily constrained by the amount of time it will take to annotate it, which further impacts the generalizability of the study. Similar challenges are faced when considering conversations from low-resource languages, which may be harder to annotate by virtue of their limited resources. This is one of the major issues we address in the present study \textit{i.e.,} can we automate the process of identifying when and how the listener backchannels from the data, thereby easing the annotation process?
\end{enumerate}

Our novel contributions include:
\begin{enumerate}
    \item Use of self-training based semi-supervision for labeling the instances in a dataset for the presence or absence of listener backchannels, and therefore, exploring computational techniques to guide the development of conversational agents. This \textit{identification} model partly replaces the human annotator, and therefore uses the listener's multimodal features to detect his/her backchannel responses. In addition to identifying backchannel opportunities, this step also identifies the type of signals (verbal, visual or both) associated with the backchannels. To analyze the feasibility of automating the labeling process via semi-supervision, we compare the backchannel \textit{prediction} models trained on the labels assigned by the semi-supervised identification process, with the models trained on the ground truth labels (from the annotators). 
    
    \item Inspired by Bevacqua \textit{et al.}'s work \cite{bevacqua2012listener} on personality contingent listener backchannels, we statistically analyzed how people with varying personalities emit different types of backchannel responses. In particular, we study the impact of the extraversion trait of a subject on their preference of modality  \cite{bevacqua2008listening} for their backchannel response.  
    
    \item Finally, unlike most prior works, in addition to predicting the backchannel opportunities, we also predict the signal for the listener agent. The signal prediction task itself is way more challenging than the opportunity prediction task as the signals may vary significantly from person to person. We approach this task by first predicting the type of signal to emit (visual, verbal, or both). Then, based on our findings in $(2)$ and depending on what personality we want our virtual listener to embody, we select the exact signal combination for it to express.  
\end{enumerate}


With this work, unlike most of the past studies that have focused on English datasets, we are also amongst the first to use peer-to-peer conversations in Hindi, which is a low-resource language, for listener backchannels\footnote{However, this does not entail that the techniques we propose cannot be used for datasets from other languages}. In particular, we use the conversational dataset collected by Khan \textit{et al.} \cite{khan2020vyaktitv}, and annotate it for analyzing backchannels. Our quantitative and subjective evaluations reveal that:

\begin{enumerate}[(i)]
    \item By leveraging semi-supervision for identification of listener backchannels, we were able to detect the presence of backchannels $\sim 90\%$ of the times, and the type of signals associated $\sim 85\%$ of the times, with only a small subset ($25\%$) of labeled data.

    
    \item Comparing the prediction models trained on the labels generated by the identification models, with those trained on manually-annotated labels: the former setting is able to reach $\sim 93\%$ of the latter's performance in case of opportunity prediction, and $\sim 96\%$ for signal category prediction. Note that the cost performance is significantly less for the former as it needs only a small amount of labeled data, thereby substantially reducing the efforts required in annotating the data.
    
    \item Subjective analysis in the form of a user study with twenty seven participants supported our quantitative observations. Approximately $75\%$ of the subjects found the backchannels produced by our proposed model more or equally natural to the responses by the model trained on annotated labels. The participants also confirmed our observations of personality impacting the preference of modality for backchannel responses.    
\end{enumerate}

The rest of this paper has been organized as follows: Section~\ref{dataset_section} discusses the dataset used in this work, the annotation process, and the initial data analysis. In Section~\ref{methodology}, we formally describe the problem statements for the identification and prediction tasks and discuss how we model them. Additionally, this section also briefly discusses the features used for modeling these tasks. Section~\ref{results_section} begins by detailing the complete experimental setup and follows with a quantitative evaluation of all the tasks. The section also discusses the observations of a user study, performed to analyze the efficacy of our models in real-time. Finally, we conclude the paper with Section~\ref{conclusion}, which discusses the future scope of this work.

\section{Dataset}
\label{dataset_section}
In this work, we use \textit{Vyaktitv}, a peer-to-peer Hindi conversations dataset, curated by Khan \textit{et al.} \cite{khan2020vyaktitv}. The dataset provides audio and video recordings of participants involved in a dyadic conversation. There are a total of $25$ conversations ($50$ individual recordings) with each one lasting $16$ minutes and $6$ seconds on an average. A total of $38$ subjects ($24$ Male, $14$ Female) were a part of the dataset. It also provides the Big Five personality traits \cite{digman1990personality} for all the subjects. Note that, to the best of our knowledge, no work has used Hindi conversations so far for analysing listener backchannels.  

 \subsection{Annotating Listener Backchannels}
 \label{annotation_details}
The audio and visual feeds for the individual speakers were annotated for verbal and visual activity based backchannels. Specifically, three annotators used the ELAN annotation software \cite{brugman2004annotating} with a custom tier template to mark the onset and offset time for different backchannel signals  including- \textit{nod}, \textit{head-shake}, \textit{mouth}, \textit{eyebrow}, and \textit{short-utterance} (Table~\ref{tab:data_desc}). The overall agreement amongst the annotators with respect to the presence of backchannels (\textit{i.e.,} if any backchannel activity was present in a particular time range)  was near perfect with a Fleiss' $\kappa$ of $0.86$. For the individual signals ($\kappa$): substantial agreement was observed for \textit{nod} ($0.71$), \textit{head-shake} ($0.64$), and \textit{mouth} ($0.63$), and moderate agreement in case of \textit{eyebrow} ($0.45$) and \textit{short-utterance} ($0.49$). Similar values were also reported by other dataset annotators, like \cite{singh2018p2pstory,Moubayed2009}. A primary reason for relatively low values of annotator agreement in such datasets is the subjective nature of backchannels.            

\begin{figure}[!t]
    \centering
    \includegraphics[width=0.95\columnwidth]{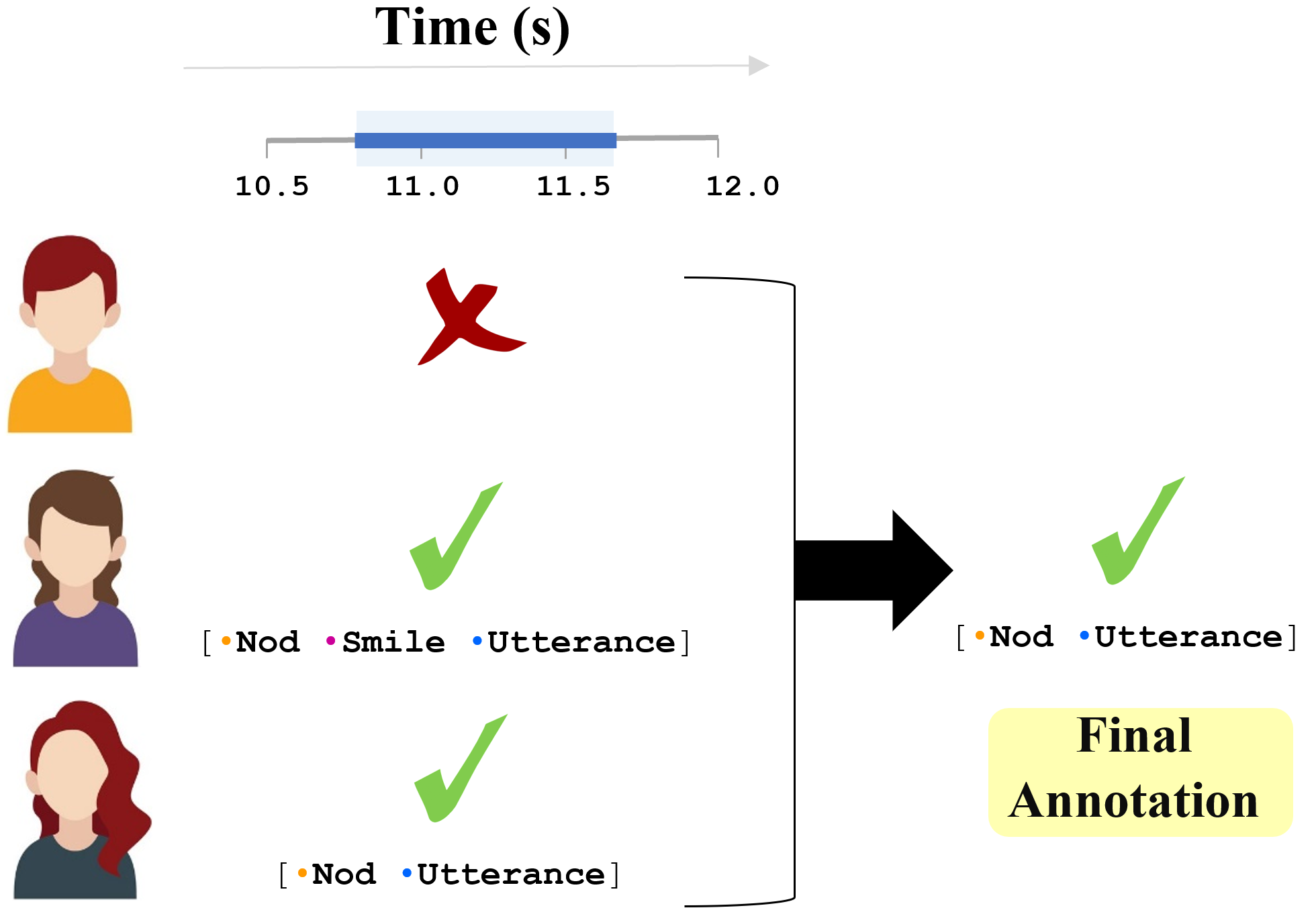}
    \caption{Sample depicting the consensus strategy adopted for combining the annotations from different coders.}
    \label{fig:annotation}
    \Description{The consensus strategy adopted for combining annotations from different coders. The figure illustrates this via an example where we have 3 annotators (A1, A2 and A3), and a particular instance (a time window of say 3 seconds). Each annotator assigns a set of annotations to this window, based on their judgement of whether different kinds of listener backchannels exist in this time frame or not. Consider a case where (i) A1 says that the instance does not consist any backchannels; (ii) A2 says its positive, and assigns the signals <Nod, Smile, Utterance> to it; and (iii) A3 also says its positive, but assigns <Nod, Utterance>. Then, combining these all, the final label for this instance would be a positive, with the signals being <Nod, Utterance>. Note that the signal Smile was annotated only by A2 (a single annotator), which is why it was not in the final label.}
\end{figure}

A consensus strategy was adopted to combine the three sets of annotations from different coders. In particular, we assigned a positive label to an instance, provided at least two annotators agreed upon the existence of backchannel activity during that time frame. The feedback signals associated with these backchannel instances (like a nod, smile, etc.) were also decided in a similar fashion using consensus voting, where a particular feedback signal is included in the final annotation if at least two annotators observed that signal's presence. Figure~\ref{fig:annotation} depicts this strategy using an example where two annotators agree upon the existence of backchannel activity.

Note that the onset and offset times annotated by the three coders were bound to have slight variation even for the same instance. Therefore, to resolve this issue while combining the annotations, any two instances from different annotation sets (labeled by different coders) whose onset and offset time were less than $1$ second apart, were considered to point to the same instance. The final value for the onset and offset time was taken as the average of the timestamps by different annotators. 

\begin{table}[!t]
\resizebox{\columnwidth}{!}{
    \centering
    \begin{tabular}{c|c|c|c|c}
    \hline
    \hline

    \textbf{BC Signal} & \textbf{Labels (\textit{Default}) } & \textbf{N} & \textbf{Mean Freq.} & \textbf{Mean Dur. (s)}\\\hline
       \rowcolor{lgrey!20} Nod & \textit{none}, nod & $2037$ & $42.4$ & $1.7$\\
        Head-shake & \textit{none}, head-shake & $207$ & $4.3$ & $1.7$\\
       \rowcolor{lgrey!20} Mouth & \textit{neutral}, smile/laugh, frown & $227$ & $4.7$ & $1.8$\\
        Eyebrow & \textit{neutral}, raise, frown & $27$ & $0.6$ & $2.1$\\
       \rowcolor{lgrey!20} Utterance & \pbox{0.4\columnwidth}{ \textit{none}, short-utterance \\ (eg: \textit{``ohh", ``okay"})} & $1161$ & $24.2$ & $1.4$\\
    \hline\hline
    \end{tabular}}
    \caption{\textbf{Backchannel Signals and their Descriptive Statistics.} (after taking consensus) \textbf{N} represents the total number of a particular feedback signal observed across the complete dataset. \textbf{Mean Freq.} is the average number of times a participant emits a particular signal during a conversation. \textbf{Mean Dur.} is the average duration of the signals in seconds (s).}
    \label{tab:data_desc}
    \end{table}

 \subsection{Data Analysis}
 \label{data_analysis}
A total of $2781$ backchannel instances were observed across all the participants in the dataset. In Table~\ref{tab:data_desc}, we present a descriptive analysis for these instances in terms of the type of feedback signals. In particular, we record the total number (\textbf{N}), average frequency per participant (\textbf{Mean Freq.}), and the average duration (\textbf{Mean Dur.}) for each signal type. Note that a backchannel instance could have multiple feedback signals (eg., nod, and smile), and therefore, has been included in each of the possible signal categories. Observations from the analysis reflect how all the participants frequently used nods and short utterances. On the other hand, backchanneling via head-shakes, mouth, and eyebrow movements was not common. In particular, instances of backchanneling through eyebrow movements were scarce in the dataset (only $27$), and therefore, we refrained from considering them in our predictive analysis. Furthermore, the average duration for all these backchanneling signals was around $1.8$ seconds.

\begin{figure}[!t]
\begin{minipage}{0.55\columnwidth}
    \centering
    \includegraphics[width=\columnwidth]{ 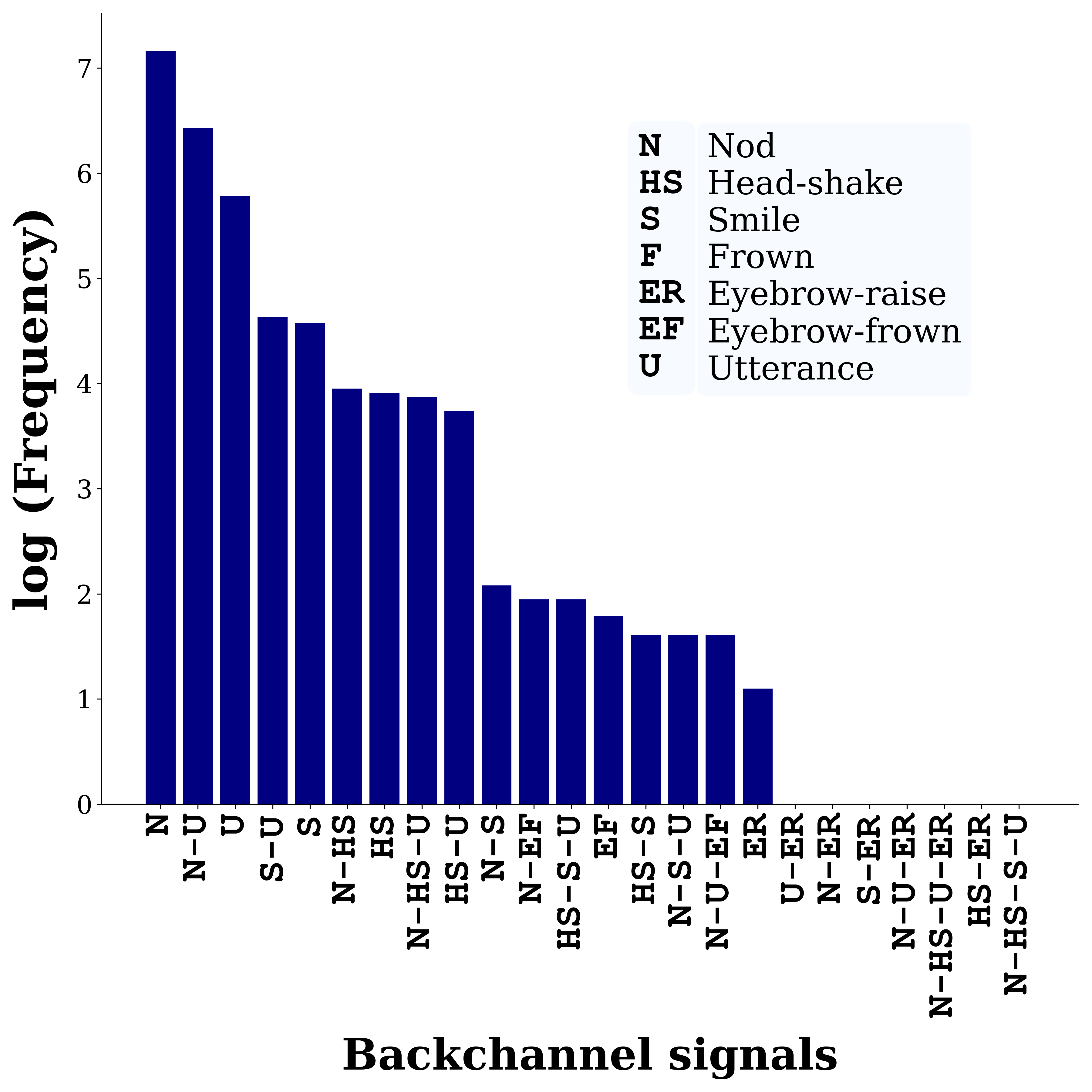}
    
    \textit{(i)}
\end{minipage}
\begin{minipage}{0.41\columnwidth}
    \centering
    \vspace{1cm}
    \includegraphics[width=\columnwidth]{ 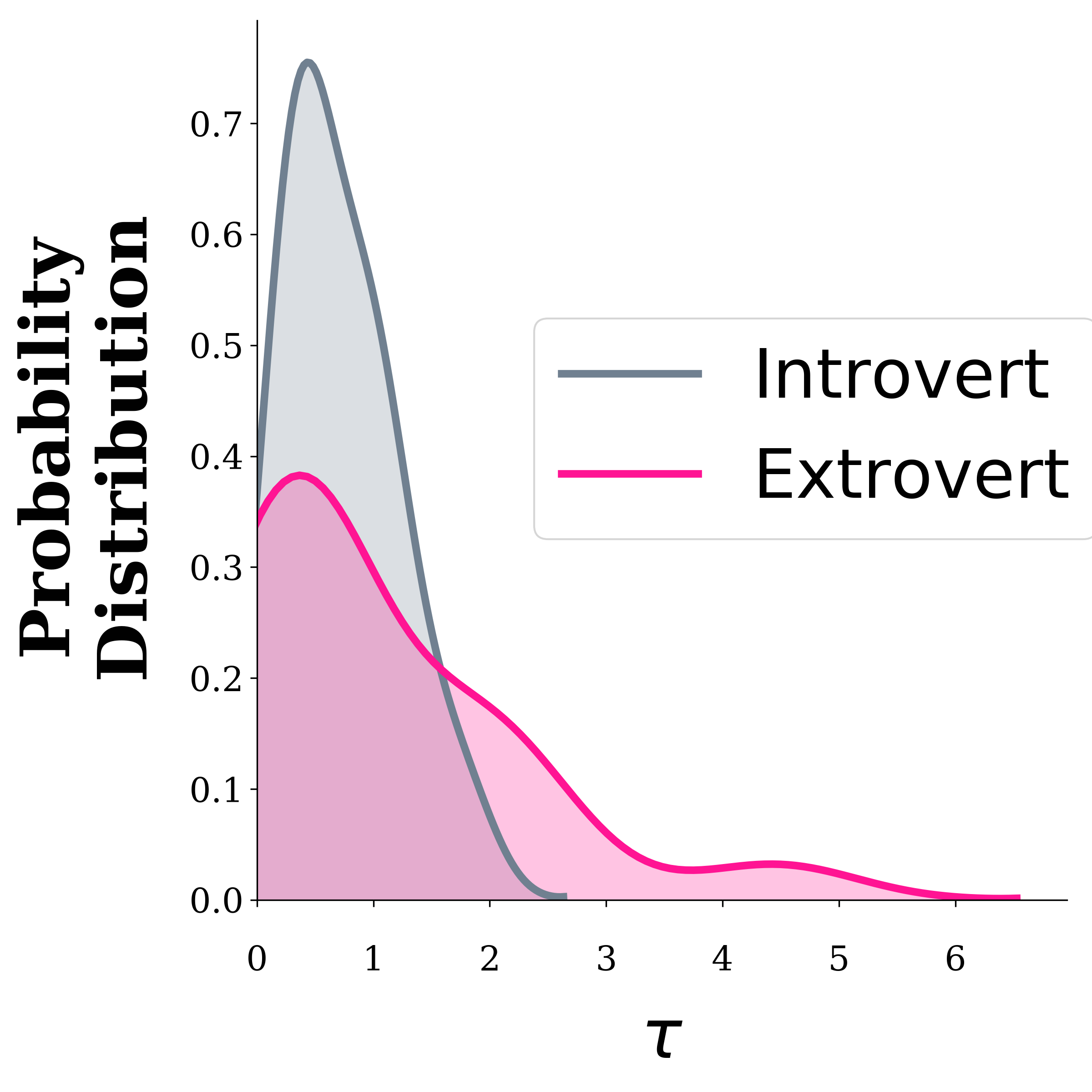}
   \\
   \vspace{5mm}
    \textit{(ii)}
\end{minipage}
    \caption{\textit{(i)} Frequency of different combinations of backchannel signals emitted together by the subjects (across all subjects). \textit{(ii)} Probability density distribution plot depicting the relation between extroversion and the ratio of  multimodal to unimodal backchannel instances (\textbf{$\tau$}) emitted by the subjects.} 
    \Description{(i) Frequency of different combinations of backchannel signals emitted together by the subjects (across all subjects). We refer to a backchannel instance as multimodal if it has multiple associated signals; else, it is a unimodal instance. Unimodal nods and utterances were amongst the top frequent signals emitted. Furthermore, most of the frequent multimodal instances included at least either a nod or an utterance. Multimodal signals where more than three types of signals co-occurred were infrequent. (ii) Probability density distribution plot depicting the relation between extroversion and the ratio of  multimodal to unimodal backchannel instances (tau) emitted by the subjects. The Kolmogorov-Smirnov test suggested a significant difference in the distribution of tau for introverts and extroverts.
}
    \label{fig:extraversion_ratio}
\end{figure}

With Figure~\ref{fig:extraversion_ratio}\textit{(i)}, we aim to dig deeper by assessing the signals that co-occur frequently. In this study, we refer to a backchannel instance as \textit{multimodal} if it has multiple associated signals; else, it is a \textit{unimodal} instance. Note how unimodal nods and utterances were amongst the top frequent signals emitted. Furthermore, most of the frequent multimodal instances included at least either a nod or an utterance. Multimodal signals where more than three types of signals co-occurred were infrequent. 


Finally, we utilize the Big Five traits provided in the dataset to analyze if personality traits influenced the type of signals (multimodal or unimodal) emitted by the subjects. For our analysis, we defined the variable $\tau$ as the ratio of the number of multimodal to unimodal backchannel instances emitted by a subject. Then, we used a two-sample Kolmogorov-Smirnov (K-S) test, with the null hypothesis that the distribution of $\tau$ for the two categories of participants (based on a particular personality trait) was similar. The test presented an interesting insight in terms of the subjects' preference for multimodal and unimodal feedbacks. Of the five traits, the tests suggested a significant difference in the distribution of $\tau$ when considering the extraversion trait. The probability density distribution in Figure~\ref{fig:extraversion_ratio} \textit{(ii)} shows how subjects with low extraversion scores (introverts) tended to have a lower value for $\tau$, indicating their preference for unimodal signals, while extroverts used multimodal feedback more often. Quantitatively, the average probabilities of an extrovert emitting multimodal or unimodal backchannels were $0.51$ and $0.49$, respectively. The same values for an introvert were $0.35$ and $0.65$, respectively. This observation has been utilized later in this study while deploying the backchannel prediction models to a virtual-agent, thereby lending it a `personality' based on which it can \textit{choose} to emit multimodal or unimodal feedbacks in a probabilistic fashion (Section~\ref{user_study_sec}).           

\subsection{Negative Samples}
\label{negative_sampling}
Negative samples are the instances from the conversations where the listener did not emit any backchannel. Specifically, we evaluated the following two conditions while extracting such instances: 

\begin{enumerate}
    \item the listener is not speaking, and
    \item the listener is not backchanneling in that time frame.
\end{enumerate}

We used the Audacity tool\footnote{\url{https://www.audacityteam.org/}} to extract the voice-activity of the listener. Combining the voice-activity and the annotations, we extracted regions from the conversations that met the above two conditions. We sampled disjoint instances from these regions, where the length of each instance (in seconds) was taken as a random floating-point number in the range [$1.06$, $5.43$]. This range also determined the lengths of the backchannel instances, and hence the choice. Using this approach, we sampled a total of $2670$ negative instances.     

\section{Methodology}
\label{methodology}
\begin{figure*}
    \centering
    \includegraphics[width=0.9\textwidth]{ 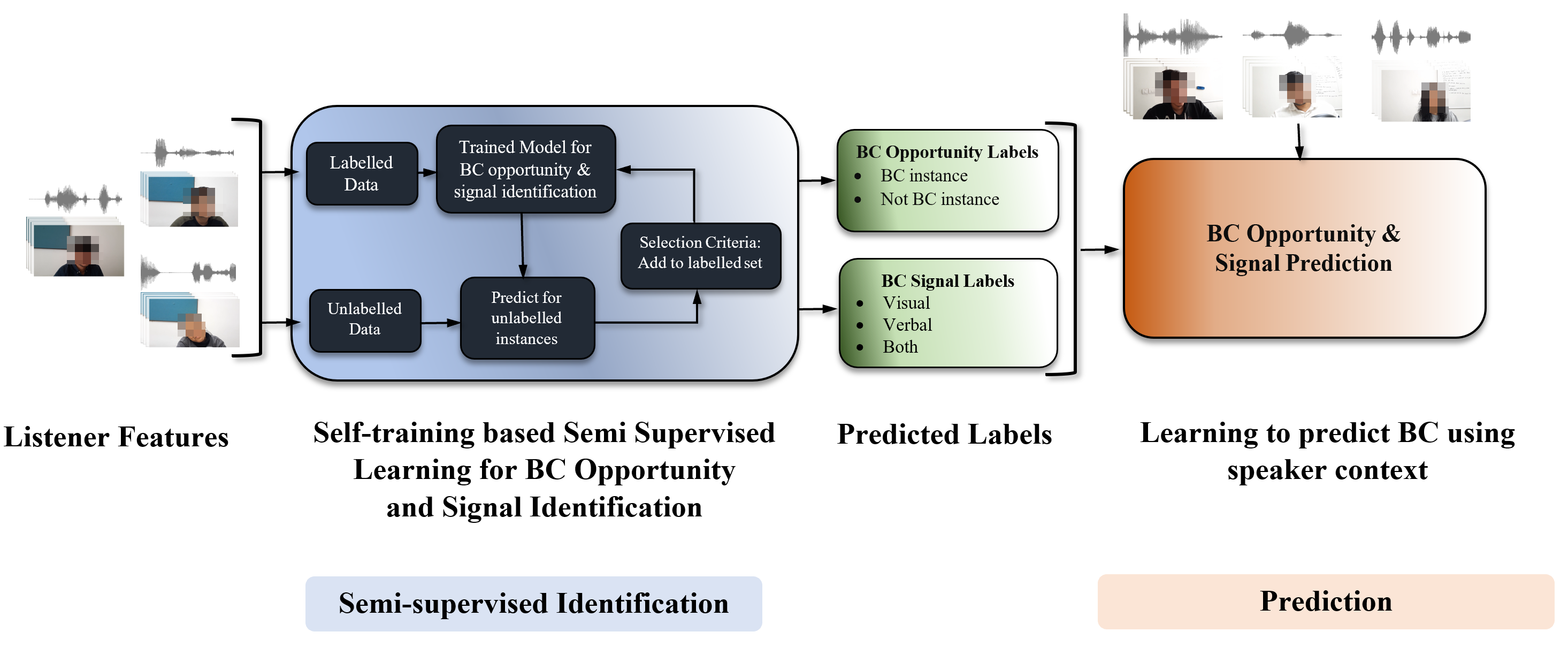}
    \caption{Methodology: \textit{(i)} Semi-supervised learning for identifying backchannels and type of signals emitted using a subset of labeled data. \textit{(ii)} Learning to predict these instances and signals using the speaker's context.}
    \label{fig:sys_arch}
    \Description{2-phase methodology: (i) Semi-supervised learning for identifying backchannels and type of signals emitted using a subset of labeled data. We use a self-training based paradigm with only a small proportion of the annotated listener backchannels, and learn a mapping to identify them based on the listener's visual and audio features. This phase intends to reduce the manual efforts involved in annotating large datasets. The pseudo-labels from this phase of identifying listener backchannels form the training labels for the next phase. (ii) Learning to predict these instances and signals using the speaker's context. We use the speaker’s visual and audio features for a context frame (a time window preceding a backchannel), along with the pseudo labels to predict these backchannels.}
\end{figure*}

This section elaborates on the two-step methodology followed in the present study. We begin with the backchannel opportunity and signal \textbf{identification} module, which utilizes the \textit{listener's features} to automatically classify the instances into different categories based on the presence or absence of backchannel activity. This step employs a semi-supervised training paradigm and aims to simplify the manual annotation task. Next, we discuss the \textbf{prediction} module, which uses the \textit{speaker's contextual} features to predict these backchannels. The training step of this latter module makes use of the labels generated by the former semi-supervised step. Figure~\ref{fig:sys_arch} summarises this workflow.

\subsection{Semi-Supervision for Backchannel Opportunity and Signal Identification}
\label{bc_identification_methodology}
 \noindent \textbf{Task Formulation.} We begin this module by first formally defining the task of identifying backchannel opportunities and signals. Consider a time frame $T_{ij}$, which starts at the $i$th second and ends at the $j$th, and let $L_{ij}$ represent the listener's visual and acoustic features in that time frame. Then, our aim for the backchannel opportunity identification task is to learn a function $\mathcal{F}_{bc}$ mapping a time series in the listener's feature space to the corresponding backchannel opportunity label $\mathcal{BO}_{ij}$ (binary label signifying the presence or absence of backchannels in the time $T_{ij}$), \textit{i.e.},

\begin{equation}
\mathcal{F}_{bc}(L_{ij}) \mapsto \mathcal{BO}_{ij} 
\end{equation}

 We model (identify and predict) the backchannel signals differently from the literature.  Instead of identifying the different signals (like nod, head-shake, etc.) individually, we categorize them into two types- \textit{visual} (nod, mouth, head-shake) and \textit{verbal} backchannels (utterances). This is done because the task of predicting the exact backchannel signal that a listener must emit based on the speaker's context (\textit{i.e.,} the signal prediction task
 ) is challenging, primarily because of the subjective nature of these signals. For instance, subject A may emit a smile to a particular speaker context, whereas another subject B may emit a nod in response to the same. Grouping signals together into visual and verbal categories simplifies the tasks at hand.

 A backchannel instance could be associated with visual, verbal, or both kinds of signals. Therefore, the signal identification task aims at finding the type of signal the listener emits whenever s/he backchannels. The goal is to learn a mapping function $\mathcal{F}_{sig}$ from the listener's feature space to one of the three signal categories (verbal, visual, both) ($\mathcal{BS}_{ij}$), \textit{i.e.,}

\begin{equation}
    \mathcal{F}_{sig}(L_{ij}) \mapsto \mathcal{BS}_{ij}
\end{equation}

It is important to note that both the identification tasks make use of only the listener's features. This is a crucial distinction from the prediction module. The identification module identifies backchannel opportunities and signals much like a human annotator, by paying attention to the listener. On the other hand, as expected, the prediction module would use only the speaker's contextual features to predict these backchannels. \vspace{2mm}\\

 \noindent \textbf{Modelling: Semi-Supervised Learning.} Annotating large datasets is a tedious task, and researchers have long been exploring ways to ease out and automate this manual process \cite{hollandi2020annotatorj, ayache2008video, rubinstein2012annotation, bota2019semi}. Even in the context of the present study, the annotation process took around $90$ hours\footnote{$30$ hours each annotator, taking the average time to annotate one side of a conversation as $35$ minutes. The total amount of recorded content being $\sim 13.5$ hours long.}, where the three annotators viewed and labeled all the conversations. This indeed is a bottleneck! To the best of our knowledge, this challenge in developing ECAs, specifically for modeling the backchannel behavior of an active listener, is an open research gap that has not been investigated by prior literature. The identification module of our workflow is a step towards tackling this challenge.

Several AI techniques based on Semi-Supervision \cite{goswami2020discriminating, hong2020x} and Weak-Supervision \cite{lison2020named, dunnmon2020cross} have been utilized to decrease the annotation costs in different application domains. Here we explore self-training based semi-supervised learning paradigm \cite{van2020survey} for identifying listener backchannel instances and the signals associated with them, using only a small subset of the manually annotated data. The following steps summarise the general approach to self-training based semi-supervision adopted here (Figure~\ref{fig:sys_arch}):
\begin{enumerate}
    \item We start with the labeled portion of the dataset ($\mathcal{L}$) to train an initial classifier ($\mathcal{C}$) that learns to identify the backchannel instances and the signals associated based on the listener's features.
    
    
    \item  $\mathcal{C}$ is then used to predict the labels for the unlabeled data ($\mathcal{U}$).
    
    
    \item Of these predictions, the instances which meet a specific \textit{selection criterion} are removed from $\mathcal{U}$ and added, along with their predicted pseudo labels, to the training set. This updated training set comprising the initially labeled ($\mathcal{L}$) and the newly added pseudo labeled instances are used to train the classifier $\mathcal{C}$ again. 
    
    \item The cycle continues until no new instance matches the selection criteria. 
\end{enumerate}

  In the present study, we have the labels for the complete dataset ($\mathcal{D}$). Therefore, in our context, $\mathcal{L}$ and $\mathcal{U}$ are disjoint subsets of $\mathcal{D}$\footnote{This notion becomes more apparent as we discuss the experimental setup in Section~\ref{experimental_setup}}. Furthermore, as selection criteria for Step-3, we use a high threshold value of $0.90$ on the predicted class probability. Note that after Step-4, we use the trained classifier $\mathcal{C}$ to predict the pseudo labels for all the remaining instances from $\mathcal{U}$. Our experiments revealed that such instances were very few in number. 

\subsection{Backchannel Opportunity and Signal Prediction}
\label{bc_prediction_methodology}
 The formal definition for the backchannel opportunity prediction task is similar to the one proposed in prior literature \cite{goswami2020social}. Consider a time window $T_{ij}$, and let $S_{ij}$ be the speaker's visual and acoustic features for that period. The backchannel opportunity prediction task entails predicting whether the listener would backchannel after $T_{ij}$ (\textit{i.e.,} the label $\mathcal{BO}_{ij+}$) using only the speaker's features (context). Similarly, the signal prediction task aims to predict the type of feedback signal (visual, verbal, or both) (\textit{i.e.,} $\mathcal{BS}_{ij+}$) that will be emitted by the listener using the speaker's context. The following function mappings represent these tasks:

\begin{equation}
        \mathcal{G}_{bc}(S_{ij}) \mapsto \mathcal{BO}_{ij+}
\end{equation}

\begin{equation}
        \mathcal{G}_{sig}(S_{ij}) \mapsto \mathcal{BS}_{ij+}
\end{equation}

Inspired by literature \cite{goswami2020social, 10.1007/978-3-319-20916-6_31} we extract a $3$ second context window before each instance in our dataset and use the speaker's visual and acoustic features for that time frame to predict listener backchannels.

The prediction tasks are performed in a supervised fashion, with the labels derived from the semi-supervision based identification module. Since the identification module used the listener's channel (features), the prediction module is essentially learning based on cross-channel semi-supervised labels. 

\subsection{Feature Extraction}
\label{featureeng}
This section elucidates all the features extracted from the dataset for modeling the identification and prediction tasks. Table~\ref{tab:features} summarises these visual and prosodic features. In particular, inspired by prior work \cite{goswami2020social, jindal2020makes} we use OpenFace \cite{baltrusaitis2018openface} to extract- $18$ facial action units (FAUs), velocity and acceleration of gaze, translational and rotational head velocities and accelerations, blink rate, pupil location, and smile ratio. Additionally, we also find the gaze state as a categorical variable, taking up three values- \texttt{left}, \texttt{right}, or \texttt{blinking}\footnote{The following GitHub repository was used for the same: \url{https://github.com/antoinelame/GazeTracking}}. These features from the listener's channel are used for backchannel opportunity, and signal \textbf{identification} tasks and the \textbf{prediction} tasks utilize the corresponding features from the speaker's channel.

As prosodic features, in addition to the voice activity (discussed in Section~\ref{negative_sampling}), we also extract the fundamental frequency (\texttt{F0}), the energy, and the first $13$ Mel-Frequency Cepstral Coefficients (MFCC) using the pyAudioAnalysis library \cite{giannakopoulos2015pyaudioanalysis}. All of them are used for the identification as well as the prediction tasks.

\begin{table}[!t]
\centering
\resizebox{0.8\columnwidth}{!}{
\begin{tabular}{P{0.28\columnwidth}p{0.71\columnwidth}}
\hline\hline
\textbf{Features} & \multicolumn{1}{c}{\textbf{Description}} \\

\rowcolor{lgrey!30} \multicolumn{2}{c}{\textbf{Visual Features}} \vspace{2mm}\\
\texttt{FAUs} &  Regressive values of 18 Facial Action Units: \texttt{AU01\_r, AU02\_r, AU04\_r, AU05\_r, AU06\_r, AU07\_r, AU09\_r, AU10\_r, AU12\_r, AU14\_r, AU15\_r, AU17\_r, AU20\_r, AU23\_r, AU25\_r, AU26\_r, AU28\_r, AU45\_r} \vspace{3mm}\\

\texttt{gaze\_vel, gaze\_acc} & Velocity and acceleration of eye gaze \vspace{3mm} \\

\texttt{gaze\_state} & Categorical feature signifying the direction of gaze as- \texttt{left}, \texttt{right}, \texttt{blinking} \vspace{3mm} \\

\texttt{head\_vel\_T}, \texttt{head\_acc\_T} & Translational velocity and acceleration of head \vspace{3mm} \\
\texttt{head\_vel\_R}, \texttt{head\_acc\_R} & Rotational velocity and acceleration of head \vspace{3mm} \\
 \texttt{blink\_rate} & First order differential of Eye Aspect Ration \vspace{3mm} \\ 
\texttt{pupil} & Location of the pupils. \vspace{3mm} \\ 
\texttt{smile\_ratio} & Stretch of the smile calculated as the ratio of two characteristic dimensions of the mouth \cite{Moubayed2009} \vspace{3mm} \\
\rowcolor{lgrey!30} \multicolumn{2}{c}{\textbf{Prosodic Features}} \vspace{2mm}\\
$\texttt{F0}$ & The fundamental frequency of the speech signal \vspace{3mm} \\
$\texttt{energy}$ & The sum of squares of the signal values, normalized by the respective frame length \vspace{3mm} \\
$\texttt{mfcc}$ & Mel-Frequency Cepstral Coefficients 1-13 \vspace{3mm} \\
\texttt{voice\_activity} & Binary state characterizing whether the person is speaking or not, based on acoustic signals. \vspace{3mm} \\
\hline\hline
\end{tabular}}
\caption{Visual and vocal prosodic features used in the study.}
\label{tab:features}
\vspace{-5mm}
\end{table}

\subsection{Experimental Setup}
\label{experimental_setup}
We now discuss the experimental settings for our work. Since each task can be analyzed and evaluated based on several dimensions, such as the type of features used (multimodal vs. unimodal), and the amount of initial `labeled' data in semi-supervision ($\mathcal{L}$), we could theoretically have had a combinatorial number of settings corresponding to all pairs of values for these variables. However,  we follow a more structured pattern to limit the number of settings and draw focus on the most crucial elements. \\

\subsubsection{Backchannel Opportunity and Signal Identification.} Separate models were trained for backchannel opportunity and signal identification tasks. We tried several different machine learning algorithms for training the classifier $\mathcal{C}$ in a semi-supervised paradigm described in Section~\ref{bc_identification_methodology}. Specifically, we experimented with Random Forests (\textbf{\texttt{RF}}), Support Vector Machine Classifier (\textbf{\texttt{SVC}}), K-Nearest Neighbour Classifier (\textbf{\texttt{KNN}}), AdaBoost (\textbf{\texttt{ADA}}), and ResNet (\textbf{\texttt{ResNet}}). ResNet is the state-of-the-art deep learning model for time series classification \cite{fawaz2019deep}. Additionally, we also experimented with the widely used Label Spreading algorithm (\textbf{\texttt{LSpread}}, implemented as a part of the scikit-learn python library \cite{scikit-learn}) for semi-supervised learning as a baseline. All but for the ResNet model were trained using the mean and standard deviation aggregates of the time series based listener features. ResNet was trained using the detailed time series features for the complete time window. Finally, for the proportion ($x$) of the dataset $\mathcal{D}$ taken as the initial `labeled' data ($\mathcal{L}$) for training $\mathcal{C}$, we experimented with all values in the range $(5\%, 100\%)$ with a step size of $5\%$. Although semi-supervision is applicable only when the amount of unlabelled data exceeds the labeled, experimenting with the whole range helps in the analyzing the models' sensitivity, \textit{i.e.}, how the performance changes when we approach the fully supervised setting (by increasing $x$).

\begin{figure}[t!]
    \centering
    \includegraphics[width=0.75\columnwidth]{ 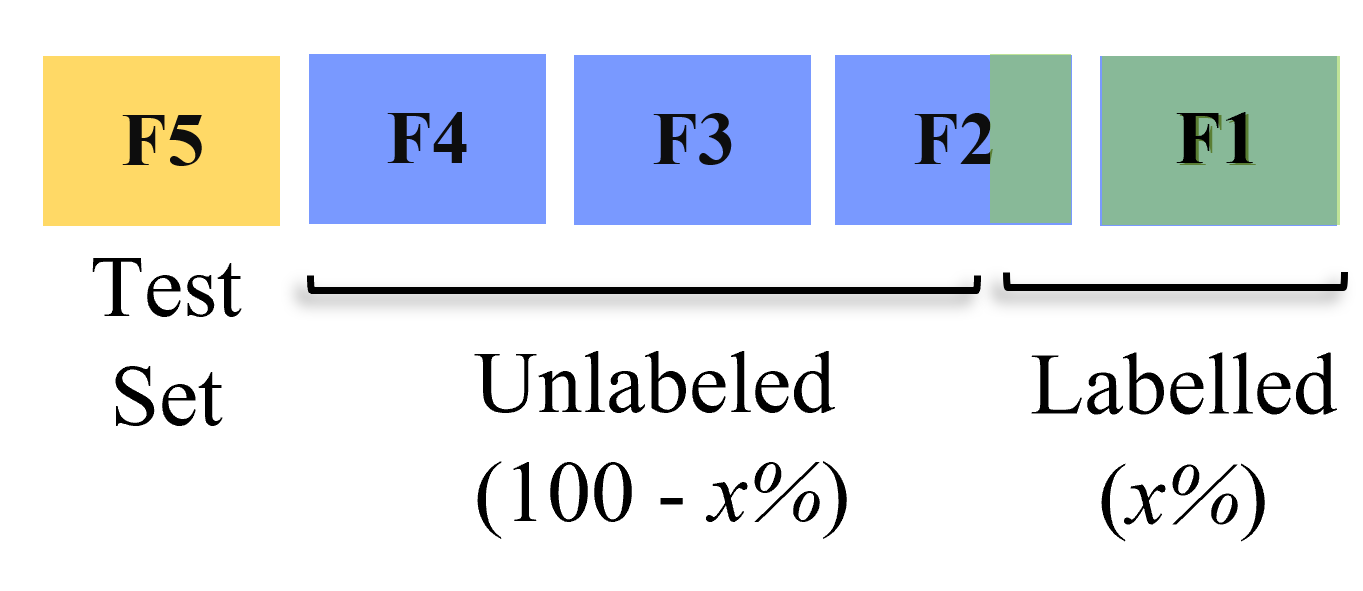}
    \caption{5-Fold evaluation of semi-supervised models for backchannel opportunity and signal identification.}
    \label{fig:ssl_splits}
    \Description{5-Fold evaluation of semi-supervised models for backchannel opportunity and signal identification. First, we randomly create 5 folds from the data (D). Four of these folds are used for training the classifier C via semi-supervision; i.e. a random x\% (the value x is parameterized) sample of the data from these four folds serve as the initial `labeled' set  (L), while the rest is termed as `unlabelled' (U). Once the model has been trained, it is evaluated on the 5th fold, i.e., the pseudo labels produced by the identification models are compared against the ground truth labels (by the annotators).}
\end{figure}

Manifold evaluation for the identification tasks can be easily understood from Figure~\ref{fig:ssl_splits}. First, we randomly create $5$ folds from the data ($\mathcal{D}$). Four of these folds are used for training $\mathcal{C}$ via semi-supervision; \textit{i.e.,} a random $x\%$ sample of the data from these four folds serve as the initial `labeled' set  ($\mathcal{L}$), while the rest is termed as `unlabelled' ($\mathcal{U}$). Once the model has been trained, it is evaluated on the $5$th fold, \textit{i.e.}, the pseudo labels produced by the identification models are compared against the ground truth labels. To ensure that our models do not over-fit to a particular random split, we run $10$ simulations of training and evaluation for each pair of values of $\mathcal{C}$ and $x$, and report the average results across all the simulations\footnote{The results for each simulation were taken as the average of all $5$ folds.}. As evaluation metrics, we use the weighted average precision, recall, F1-score, and overall accuracy, for both the opportunity and signal identification tasks.

\hfill
\begin{figure}[t!]
        \centering
    \includegraphics[width=\columnwidth]{ 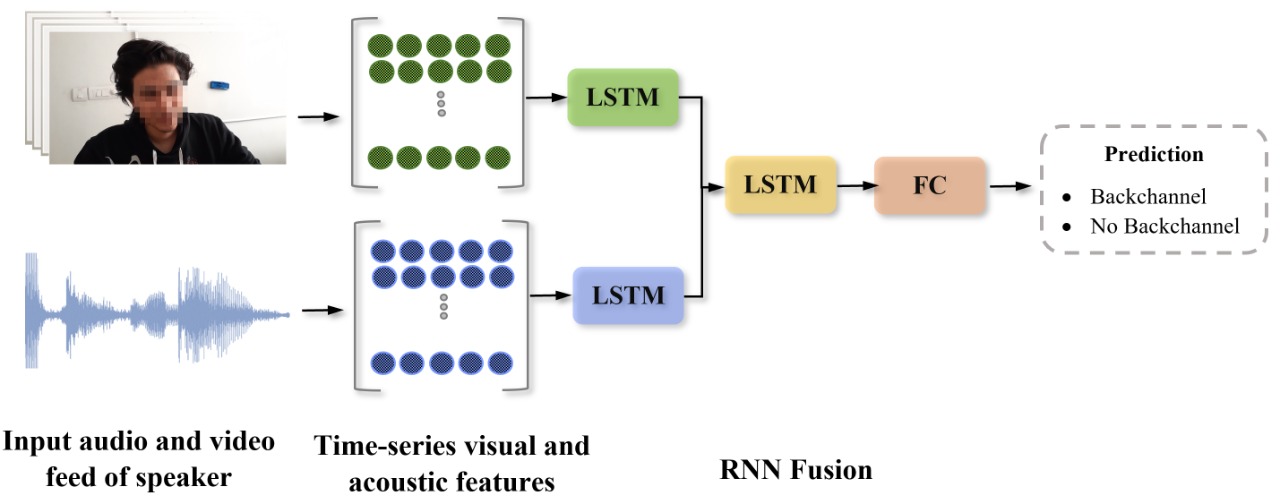}
    \caption{Multimodal RNN Fusion based architecture for backchannel opportunity prediction model.}
    \label{fig:bop_pred_model}
    \Description{Multimodal RNN Fusion based architecture for backchannel opportunity prediction model. In essence, the time-series features for video and audio modalities are first individually passed through LSTM encoders. Following this, the encoders' outputs are concatenated and again passed through a recurrent layer, enabling the model to learn the temporal dependencies between modalities. The latent representation from this fusion layer is finally passed to a softmax layer through a fully connected layer to predict feedback opportunities. The two unimodal models simply include a softmax layer after the encoder to output predictions.}
\end{figure}

\subsubsection{Backchannel Opportunity and Signal Prediction.} We begin evaluating the prediction experiments by first analyzing the impact of using different sets of features \textit{i.e.,} unimodal vs. multimodal features. This is done by using the supervised (manually annotated) labels for both training and evaluation. Inspired by Tavabi \textit{et al.} \cite{multimodalempathy2019rnn}, we use a multimodal RNN fusion-based architecture for backchannel opportunity prediction, shown in Figure~\ref{fig:bop_pred_model}. In essence, the time-series features for video and audio modalities are first individually passed through LSTM encoders. Following this, the encoders' outputs are concatenated and again passed through a recurrent layer, enabling the model to learn the temporal dependencies between modalities. The latent representation from this fusion layer is finally passed to a softmax layer through a fully-connected layer to predict feedback opportunities. The two unimodal models simply include a softmax layer after the encoder to output predictions.

The signal prediction task was more complex, as the labels for the task also suffered from a serious class imbalance (\textit{visual:} $1593$, \textit{verbal:} $326$, \textit{both}: $835$) which was adversely impacting the predictive performance of the models\footnote{Note that though the same data was also used for the signal identification task, the imbalance did not cause a degradation in the performance there.}. Therefore, we use the mean and standard deviation aggregates of the time series based speaker context features, along with \texttt{SVM-SMOTE} \cite{smote} to handle this imbalance\footnote{SMOTE takes data of the shape ($n\_samples$, $n\_features$) as input, and therefore, time-series data had to be aggregated. This also eliminated the utility of models like ResNet for the task.}. Note that signal prediction was made using only the feature set (unimodal/multimodal) that performed the best for the opportunity prediction task. Using the upsampled data\footnote{Only the training set was upsampled i.e., the test set remained as is.}, we tried several models for signal prediction, including Random Forests (\textbf{\texttt{RF}}), Support Vector Machine Classifier (\textbf{\texttt{SVC}}), AdaBoost (\textbf{\texttt{ADA}}), K-Nearest Neighbors (\textbf{\texttt{KNN}}), and a Multi-Layered Perceptron model (\textbf{\texttt{MLP}})\footnote{Please refer the supplementary material for details on the hyper-parameters used for all the different models in the identification and prediction tasks }.

The next set of experiments analyze the feasibility and the performance to cost trade-off when adopting semi-supervision to label the dataset. For this, we compare the prediction tasks in two paradigms: 

\begin{enumerate}
    \item A complete supervised setting, using true labels provided by the annotators during the training and evaluation phases. Note that previous evaluations of the prediction tasks already produced the results for this setting.
    
    \item The proposed setting using labels generated by the semi-supervised identification modules for training, while evaluating on the true labels. For this, we use only the best feature set (unimodal/multimodal) found via the previous evaluations, along with the pseudo labels produced by the best-found pair of $\mathcal{C}$ and $x$ from the identification tasks.
    
\end{enumerate}

With these two paradigms, we assessed how far semi-supervised backchannel identification fetches in comparison to human annotations for the downstream prediction tasks.

For evaluating the backchannel prediction models, several recent works have used the leave-one-subject-out approach \cite{jindal2020makes}. In the present study, we use a slightly modified version of this technique to evaluate our prediction models in different paradigms. Instead of having the data from each subject comprise a test fold, we divide the subjects into $6$ groups, and data from each of these groups form a test fold, \textit{i.e.,} leave-one-group-out. This modification was primarily done because some subjects had very few backchannel instances in their conversations. Combining instances from multiple subjects to form a test set helped in better analyzing the model predictions and keeping the number of folds tractable. As metrics for the backchannel opportunity prediction task, we report the positive class' precision, recall, and F1-score, as well as the overall accuracy. For signal prediction, we report the weighted average precision, recall, and F1-score, the overall accuracy, and the confusion matrices.

\section{Results}
\label{results_section}
\subsection{Identification Tasks}

\begin{figure}[!th]
\begin{minipage}{\columnwidth}
    \centering
    \includegraphics[width=\columnwidth]{ 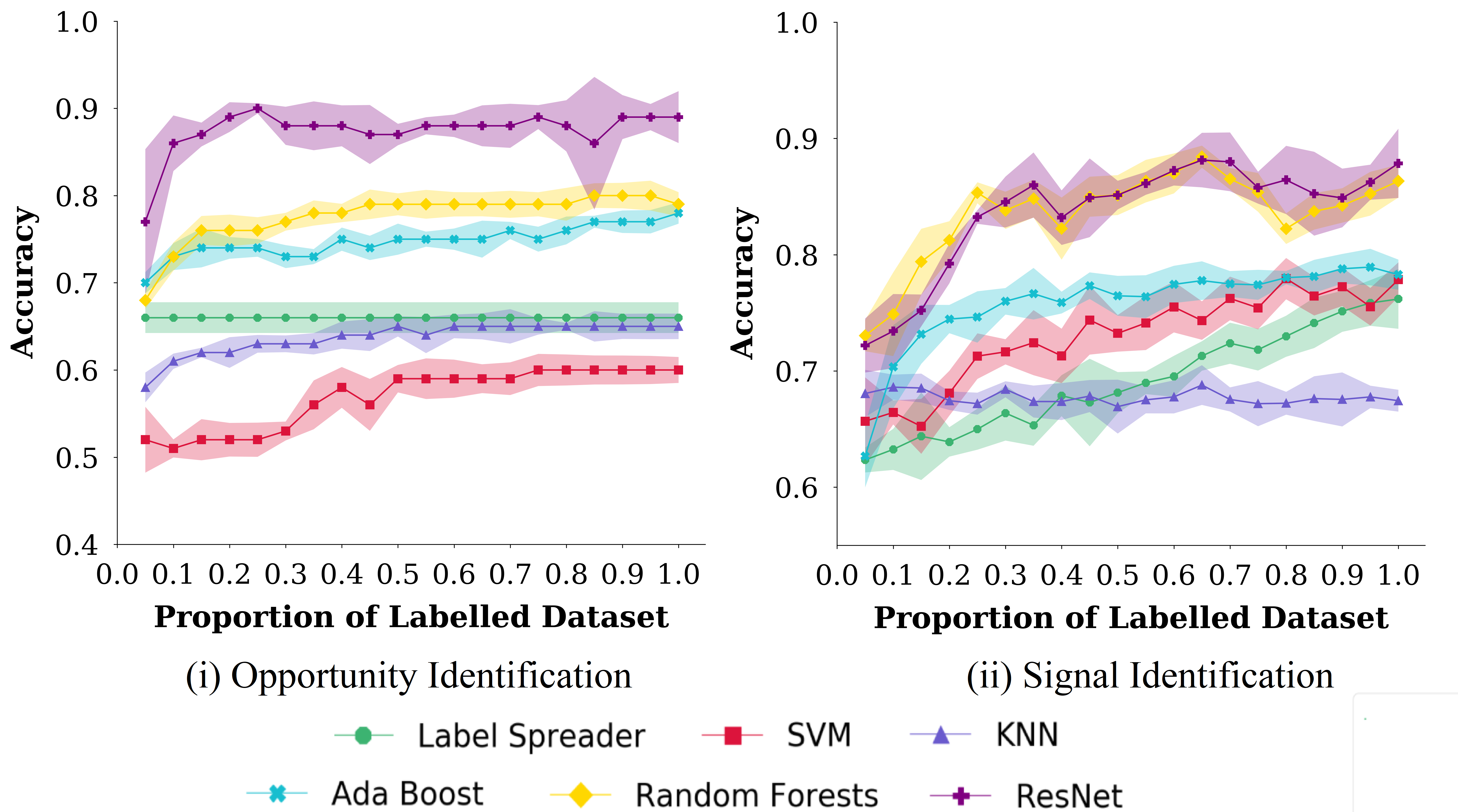}
\end{minipage}\\
    \caption{Sensitivity Analysis using accuracy as the metric for semi-supervised backchannel opportunity and signal identification models. The shaded portion represents standard deviation.}
    \label{fig:sensitivity_analysis}
    \Description{Sensitivity Analysis using accuracy as the metric for semi-supervised backchannel (i) opportunity and (ii) signal identification models. The shaded portion represents standard deviation. In essence, we plot the accuracies for each model at different values of x- the amount of seed data available while training the semi-supervised identification model. This x varies from 5\%-100\% in intervals of 5\%.
}
\end{figure}

In Figure~\ref{fig:sensitivity_analysis}, we have the sensitivity analysis in terms of accuracy of all the models we tried for the backchannel opportunity and signal identification tasks. Each curve in these plots represents how the performance of a particular classifier ($\mathcal{C}$) for the identification task changes as we increase the amount of initial labeled data available ($x$) for semi-supervision. Using this information, we looked for the best possible pair of values for $\mathcal{C}$ and $x$. However, in addition to achieving high performance, we also wanted to limit $x$. For this, we manually found the \textit{elbow} point for the best performing classifier in each task. The elbow value for $x$ is one where decreasing it further would cause a drastic drop in performance, while increasing it would not change the performance significantly. Ideally, we also wanted $x$ to be less than $50\%$, where we could have more unlabelled than labeled data for semi-supervision, thereby improving the annotation cost trade-off. 

\begin{table}[!th]
    \centering
    \resizebox{\columnwidth}{!}{
    \begin{tabular}{c|cccc|cccc}
    \hline\hline
\multirow{2}{*}{\textbf{Model}} & \multicolumn{4}{c|}{ \textbf{\footnotesize{ \makecell{Opportunity \\ Identification}} }} & \multicolumn{4}{c}{\small \textbf{\makecell{Signal \\ Identification}}}\\ 
         & \textbf{\texttt{\small Precision}} & \textbf{\texttt{\small Recall}} & \textbf{\texttt{\small F1-score}} & \textbf{\texttt{\small Accuracy}} & \textbf{\texttt{\small Precision}} & \textbf{\texttt{\small Recall}} & \textbf{\texttt{\small F1-score}} & \textbf{\texttt{\small Accuracy}}\\\hline
      \textbf{\texttt{LSpread}}  & $0.66$ &	$0.66$ & $0.66$ &	$0.66$
 & $0.67$ & $0.65$ & $0.66$ & $0.65$ \\
      \textbf{\texttt{SVC}}   & $0.56$ & $0.52$ & $0.56$ & $0.54$
 & $0.69$ & $0.70$ & $0.68$ & $0.71$\\
      \textbf{\texttt{KNN}}  & $0.63$ &	$0.63$ & $0.63$ & $0.63$
& $0.63$ & $0.67$ &	$0.62$ & $0.67$\\
      \textbf{\texttt{ADA}}  & $0.74$ &	$0.74$ & $0.74$ & $0.74$
 & $0.75$ & $0.75$ & $0.69$ &	$0.75$\\
      \textbf{\texttt{RF}}  & $0.76$ & $0.76$ & $0.76$ & $0.76$
 & \cellcolor{lgrey!30} $\textbf{0.86}$ & \cellcolor{lgrey!30} $\textbf{0.87}$ & \cellcolor{lgrey!30} $\textbf{0.86}$ & \cellcolor{lgrey!30} $\textbf{0.85}$\\
      \textbf{\texttt{ResNet}}  & \cellcolor{lgrey!30} $\textbf{0.90}$ & \cellcolor{lgrey!30} $\textbf{0.90}$ & \cellcolor{lgrey!30}	$\textbf{0.92}$ & \cellcolor{lgrey!30} $\textbf{0.90}$ & $0.81$ & $0.84$ & $0.83$ & $0.83$\\
     \hline\hline
    \end{tabular}}
    \caption{Backchannel opportunity and signal identification: Detailed results for $x$ = $25\%$.}
    \label{tab:identification_results_25}
\end{table}

Figure~\ref{fig:sensitivity_analysis} $(i)$ shows how ResNet outperforms all the other models in terms of the listener backchannel opportunity identification task. Observe how increasing $x$ beyond $25\%$ does not change the performance a lot. It also meets all the other criteria we discussed above. Therefore, we choose $x = 25\%$ as the final seed value (used to initialize the amount of labeled data) for the opportunity identification task. For the signal identification task (Figure~\ref{fig:sensitivity_analysis} $(ii)$), we observe that ResNet and Random Forests both have somewhat overlapping accuracy for different values of $x$. However, by taking a closer look, we analyze that for values of $x < 30\%$, Random Forests lead ResNet. The elbow point for Random Forests is at $25\%$, while that for ResNet (with a similar accuracy value) is observed at $35\%$. Therefore, for the signal identification task, our final choice is Random Forests with $x = 25\%$. The detailed results obtained for other classifiers with $x$ set to $25\%$ for both the opportunity and signal identification tasks are shown in Table~\ref{tab:identification_results_25}.

\subsection{Prediction Tasks}

We begin discussing the prediction models with Table~\ref{tab:bop_multimodal}, which records the results of our backchannel opportunity prediction task performed using the annotated labels for training, while experimenting with different subsets of input features. The unimodal-audio model, with positive class F1 of $0.74$ and overall accuracy of $0.70$, outperforms the unimodal-video model, with the corresponding metric values of $0.71$ and $0.66$, respectively. The multimodal model utilizing both the video and audio features beats the audio model by a small margin, attaining an F1 of $0.75$ and $0.72$ accuracy. Since this set has overall the best performance for the baseline opportunity prediction task with supervised labels, we use the same to train and analyze models for subsequent tasks as well.    

\begin{table}[t!]
    \centering
            \centering
            \resizebox{\columnwidth}{!}{
    \begin{tabular}{c|cccc}
    \hline\hline
 \textbf{\texttt{Feature set}}    &  \textbf{\texttt{\small Precision}} & \textbf{\texttt{\small Recall}} & \textbf{\texttt{\small F1-score}} & \textbf{\texttt{\small Accuracy}}\\\hline
    \textbf{\texttt{Video}} & $0.61$	& $0.86$ & $0.71$& $0.66$ \\ 
    \textbf{\texttt{Audio}} & $0.64$	& $0.90$ &	$0.74$ & $0.70$ \\
    \rowcolor{lgrey!30}
    \textbf{\texttt{Video + Audio}} & $\textbf{0.66}$ & $\textbf{0.89}$ & $\textbf{0.75}$ & $\textbf{0.72}$ \\
    \hline\hline
    \end{tabular}}
    \caption{Backchannel opportunity prediction model trained trained across different sets of features using supervised (manually-annotated) labels for training. Overall, the multimodal feature set performed the best.}
    \label{tab:bop_multimodal}
    \end{table}

\begin{table}[t!]
    \centering
    \resizebox{\columnwidth}{!}{
    \begin{tabular}{c|cccc}
    \hline\hline
      \textbf{Labels used}  & \textbf{\texttt{\small Precision}} & \textbf{\texttt{\small Recall}} & \textbf{\texttt{\small F1-score}} & \textbf{\texttt{\small Accuracy}}\\\hline
        \textbf{Supervised} & $0.66$ & $0.89$ & $0.75$ & $0.72$\\
        \textbf{Semi-Supervised} & $0.62$ & $0.82$ &	$0.70$ & $0.66$\\
        \hline\hline
    \end{tabular}}
    \caption{Backchannel opportunity prediction: comparison of model trained using supervised manually-annotated labels with the one using labels generated by the identification module.}
    \label{tab:bop_ssl}
    \vspace{-5mm}
\end{table}

\begin{table}[t!]
    \centering
    \resizebox{\columnwidth}{!}{
    \begin{tabular}{c|cccc|cccc}
    \hline\hline
\multirow{2}{*}{\textbf{Model}} & \multicolumn{4}{c|}{ \textbf{\makecell{Supervised \\ Labels}} } & \multicolumn{4}{c}{\textbf{\makecell{Semi-Supervised \\ Labels}}}\\ 
         & \textbf{\texttt{\small Precision}} & \textbf{\texttt{\small Recall}} & \textbf{\texttt{\small F1-score}} & \textbf{\texttt{\small Accuracy}} & \textbf{\texttt{\small Precision}} & \textbf{\texttt{\small Recall}} & \textbf{\texttt{\small F1-score}} & \textbf{\texttt{\small Accuracy}}\\\hline
      \textbf{\texttt{SVC}}  & $0.80$ & $0.80$ & $0.80$ & $0.80$ & $0.75$ & $0.75$ & $0.75$ & $0.75$ \\
      \textbf{\texttt{KNN}}  & $0.74$ & $0.74$ & $0.73$ & $0.74$ & $0.72$ & $0.72$ & $0.71$ & $0.72$ \\
      \rowcolor{lgrey!30}
      \textbf{\texttt{ADA}}  & $0.81$  & $0.81$ & $\textbf{0.81}$ & $0.81$ & $0.78$ & $0.78$ & $\textbf{0.78}$ & $0.78$ \\
      \textbf{\texttt{RF}}  & $0.78$ &	$0.78$ &	$0.78$ & $0.78$ & $0.76$ & $0.75$ & $0.75$ & $0.75$ \\
      \rowcolor{lgrey!30}
      \textbf{\texttt{MLP}} & $0.80$ & $0.81$	 & $0.80$ & $\textbf{0.87}$ & $0.75$ & $0.74$ & $0.74$	& $\textbf{0.84}$ \\
     \hline\hline
    \end{tabular}}
    \caption{Backchannel signal prediction: comparison of models trained using supervised (manually-annotated) labels, and those using labels generated by the semi-supervised signal identification Random Forest model (with $25\%$ initial labeled data).}
    \label{tab:bsp_results}
    \end{table}

\begin{table}
\centering
\begin{minipage}{0.47\columnwidth}
\centering
\resizebox{\columnwidth}{!}{
        \begin{tabular}{ccccc}
    & & \multicolumn{3}{c}{\textsc{\textbf{Predicted}}}\\
 & & \texttt{C1} & \texttt{C2} & \texttt{C3}\\ 
    \parbox[t]{0mm}{\multirow{3}{*}{\rotatebox[origin=c]{90}{\textsc{\textbf{True}}}}}  & \texttt{C1}  & \cellcolor{mred!71} $0.71$ & \cellcolor{mred!7} $0.07$ & \cellcolor{mred!23} $0.23$\\
      & \texttt{C2}  & \cellcolor{mred!7} $0.07$ & \cellcolor{mred!87} $0.87$ & \cellcolor{mred!6} $0.06$\\
      & \texttt{C3}  & \cellcolor{mred!27} $0.27$ & \cellcolor{mred!5} $0.05$ & \cellcolor{mred!67} $0.67$\\
    \end{tabular}}\\
    \vspace{2mm}
    $(i)$
\end{minipage}
\hfill
\begin{minipage}{0.47\columnwidth}
\centering
\resizebox{\columnwidth}{!}{
        \begin{tabular}{ccccc}
    & & \multicolumn{3}{c}{\textsc{\textbf{Predicted}}}\\
 & & \texttt{C1} & \texttt{C2} & \texttt{C3}\\ 
    \parbox[t]{0mm}{\multirow{3}{*}{\rotatebox[origin=c]{90}{\textsc{\textbf{True}}}}} & \texttt{C1}  & \cellcolor{mred!70} $0.70$ & \cellcolor{mred!9} $0.09$ & \cellcolor{mred!21} $0.21$\\
     &  \texttt{C2}  & \cellcolor{mred!8} $0.08$ & \cellcolor{mred!80} $0.80$ & \cellcolor{mred!12} $0.12$\\
    &   \texttt{C3}  & \cellcolor{mred!23} $0.23$ & \cellcolor{mred!6} $0.06$ & \cellcolor{mred!72} $0.72$\\
    \end{tabular}}\\
    \vspace{2mm}
    $(ii)$
\end{minipage}   
    \caption{Confusion matrices for the best backchannel signal prediction model $(i)$ trained on manually-annotated labels, $(ii)$ trained using labels generated by the signal identification models. Here, \texttt{C1}, \texttt{C2}, \texttt{C3} refer to the `visual', `verbal', and `both' class labels, respectively.}
    \label{tab:conf_matrices}
\vspace{-5mm}
\end{table}

Continuing with opportunity prediction, we now discuss the results obtained using the semi-supervised labels generated by the opportunity identification model for training (the evaluation/test set used the same supervised labels). Table~\ref{tab:bop_ssl} records these results. Evidently, with the labels generated using just $25\%$ of the annotated data (in the identification step), we are able to achieve nearly $93\%$ of the supervised F1 and $92\%$ of the supervised accuracy scores on the opportunity prediction tasks. The values of these metrics observed here are $0.70$ and $0.66$, respectively.

Moving on to backchannel signal prediction for determining the category of signals for the listener to emit, Table~\ref{tab:bsp_results} shows the results obtained for all the models trained in different settings. When using the manually-annotated labels for training, we observe that the \textbf{\texttt{MLP}} model has the best accuracy of $0.87$, however in terms of the F1-Score, the \textbf{\texttt{ADA}} model outperforms all the others. Given the data imbalance in the signal prediction task, we use F1-score as the deciding metric, and choose \textbf{\texttt{ADA}} as the best performing model, with an F1-score of $0.81$. We observe a similar performance trend when using the semi-supervised signal identification model to generate the training labels. Here as well, \textbf{\texttt{ADA}} is the best performing model in terms of F1-score, with a value of $0.78$, which is $96\%$ of the corresponding metric for the supervised setting. In Table~\ref{tab:conf_matrices}, we report the confusion matrices for the best (\textbf{\texttt{ADA}}) signal prediction models found for the two settings. We used the worst-case confusion matrix \cite{leekha2019you} computed using all the matrices generated across manifold evaluation of the model. The performance for the `visual' (\texttt{C1}) class remained almost unchanged in the two settings. The false positives between the `visual' and the `both' (\texttt{C3}) classes reduced when using semi-supervised labels. The `verbal' (\texttt{C2}) class performance dropped slightly. On the other hand, we performed better for the class \texttt{C3}, but that came at the cost of a slight increase in the false positives from \texttt{C2}. Overall, the performance of the model trained using semi-supervised labels was indeed comparable with the one trained on manually-annotated labels.

\subsection{Subjective Evaluation: User Case Study}
\label{user_study_sec}
For problems like predicting listener backchannels, quantitative evaluations based on different metrics may not be sufficient to analyze the models' efficacy. For such cases, qualitative assessments become extremely important. Many prior studies have used ECAs or robots to deploy their models and assess the backchannel predictions. However, in the present study, we follow a slightly different approach, which is partly inspired by \cite{goswami2020social} and \cite{10.1007/978-3-319-20916-6_31}. Specifically, using Apple's Memoji feature, we created two virtual avatars- \textit{Arjun} (introvert) and \textit{Karan}\footnote{The names used are hypothetical, and do not compromise anonymity.} (extrovert). We used the same Memoji technology to record (as short clips) the neutral state and different combinations of backchannel signals, mentioned in Table~\ref{tab:user_study_signals}, for each of them. Furthermore, we prepared a short video compilation (nearly $4$ minutes long) of a few speakers from the dataset used in this work. Using our prediction models\footnote{Backchannel opportunity prediction was done at an interval of $3$ seconds. Depending on the output, the signal categories were predicted. We also ensured that the models used for different speakers in the compilation were the ones where the speaker belonged to the test set.}, along with a personality contingent signal combination sampling technique, we recorded the backchanneling responses that Arjun and Karan would emit for the speakers in the compilation. Although the sampling is not the main focus of this work, we briefly describe below the steps followed for re-use by future works:

\begin{itemize}
    \item After getting the predictions for backchannel opportunity, and the signal category from our models, we use inverse transform sampling to decide whether to emit unimodal or multimodal signals (in case of visual and verbal categories only) based on the personality being modeled (extrovert or introvert). Note that the probabilities used here are calculated from the data and are mentioned in Section~\ref{data_analysis}.
    
    \item Now, we choose the exact signals to emit from Table~\ref{tab:user_study_signals} using inverse transform sampling, where the probabilities for each signal combination are calculated from the data\footnote{The only step that is contingent on personality is deciding on the unimodal vs. multimodal signals. The second step of signal sampling uses common probabilities for both characters.}.
\end{itemize}

We recorded six response videos, three for each Arjun and Karan, using the following different prediction policies:
\begin{enumerate}
    \item Random Prediction Model (\textbf{\texttt{Random}} policy): A baseline model that predicted backchannel opportunities and the signal categories at \textbf{random}.
    \item Supervised Model (\textbf{\texttt{MA}} policy): The best opportunity (multimodal) and signal category (\textbf{\texttt{ADA}}) prediction models trained on the \textbf{m}anually-\textbf{a}nnotated labels.
    \item Semi-supervised Model (\textbf{\texttt{SSL}} policy): The opportunity (multimodal) and signal category (\textbf{\texttt{ADA}}) prediction models trained with the labels generated from the \textbf{s}emi-\textbf{s}upervised identification models (\textbf{\texttt{ResNet}} and \textbf{\texttt{RF}}, respectively with $25\%$ data) used in this work.
\end{enumerate}

With these short clips recording the backchannel responses as predicted by different models for Arjun and Karan, we wanted to test the following three hypotheses:
\begin{enumerate}[start=1,label={\texttt{\textbf{\small [H\arabic*]:}}}, leftmargin=*]
    \item As perceived by a human watching the response videos, the backchannels predicted by our models look more natural, and therefore, better than those predicted by the random policy model.
    \item From a user's perspective, there is no significant difference in the quality of backchannel responses emitted by models in the settings $2$ and $3$ above, \textit{i.e.,} using semi-supervised learning with a small subset of labelled data does not impact the quality of backchannel responses generated by the final prediction models. 
    \item Finally, to assess the extent to which backchannel responses depend on personality, and the utility of using such personality contingent signal sampling: A human can judge the extraversion personality trait for Arjun and Karan based on their response videos.
\end{enumerate}

\begin{table}[t!]
    \centering
    \resizebox{\columnwidth}{!}{
    \begin{tabular}{c|c|c}
    \hline\hline
    & \textbf{Unimodal} & \textbf{Multimodal}\\\hline
    \textbf{Visual} & \makecell{nod ($0.83$), head-shake ($0.08$),\\ smile ($0.09$)} & nod + smile ($0.60$), head-shake + smile ($0.40$)\\\hline
    \textbf{Verbal} & \makecell{utter: top short utterances like \textit{okay}, \\ \textit{hmm}, \textit{haan} (yes), etc., \\ were proabbilistically sampled} &\multirow{2}{*}{\makecell{ nod+utter ($0.80$), smile+utter ($0.13$),\\ head-shake+utter ($0.05$), \\head-shake+utter+smile ($0.01$), \\nod+utter+smile ($0.01$)}}\\\cline{1-2}
    \multirow{3}{*}{\textbf{Both}} & \multirow{3}{*}{-} & \\
    & & \\
    & & \\
    \hline\hline
    \end{tabular}}
    \caption{Signal combinations from different categories used in the present study. The corresponding normalized probabilities inferred from the data and used while sampling the signal combinations are also shown.}
    \label{tab:user_study_signals}
\end{table}

\begin{figure}[t!]
    \centering
    \begin{minipage}{0.45\columnwidth}
    \centering
            \includegraphics[width=0.96\columnwidth]{ 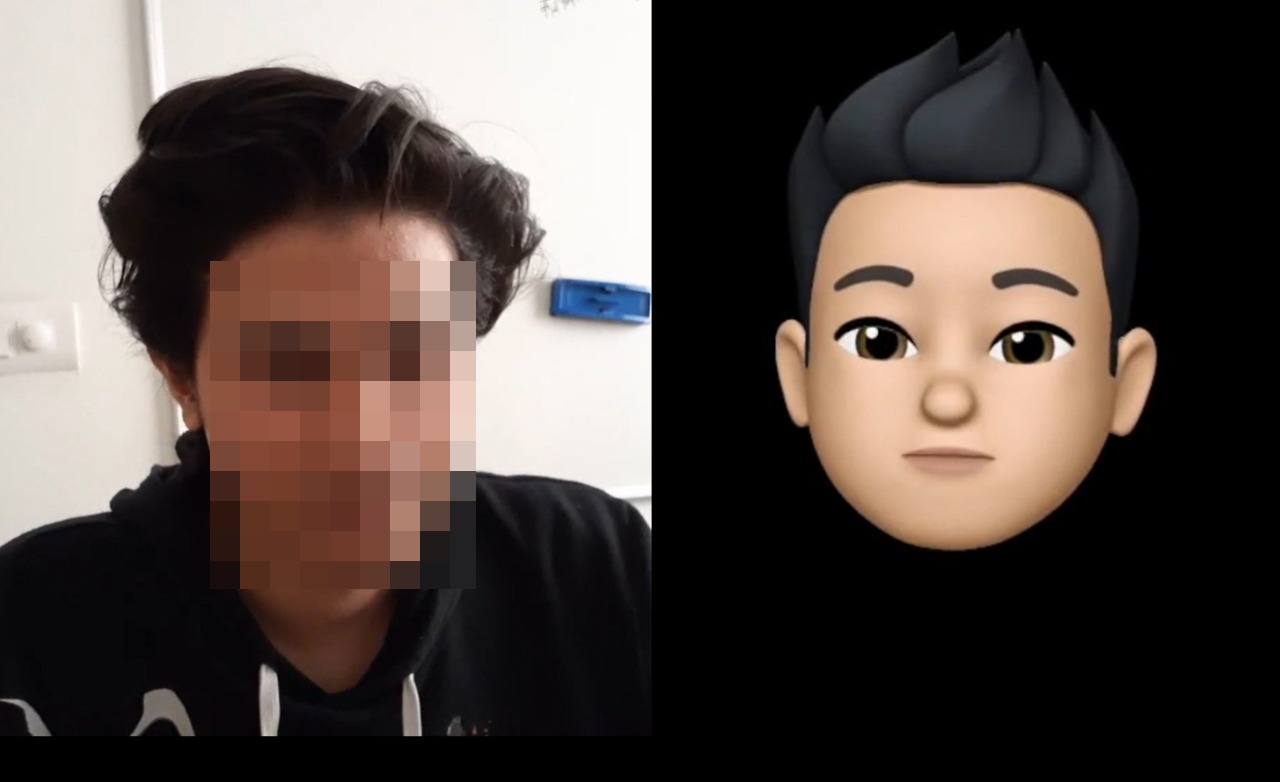}\\
            $(i)$
    \end{minipage}
        \begin{minipage}{0.47\columnwidth}
    \centering
            \includegraphics[width=\columnwidth]{ 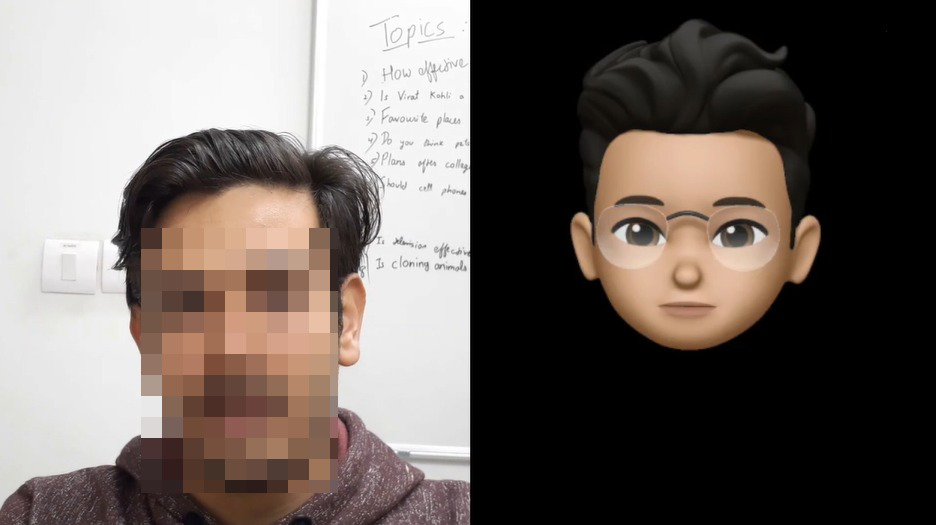}\\
            $(ii)$
    \end{minipage}
    \caption{Stills from two of our user study videos- $(i)$ Karan, and $(ii)$ Arjun}
    \label{fig:userstudy_samples}
\end{figure}

Twenty seven students were recruited as participants from a university's mailing list. They were all native Hindi speakers. As an introduction, the participants were acquainted with the concept of listener backchannels. They were instructed to watch all six response videos and carefully observe the virtual listeners' backchannel responses in each one. It was a blind study, \textit{i.e.} the participants knew neither about the prediction policies nor about the listeners' personality traits in the videos. 

Furthermore, to ensure that the appearance of the two avatars did not bias the participants' judgement, we switched the avatars (with everything else as is) for half the participants. In other words, for half of the participants, Arjun acted as an extrovert, and vice versa. With this, we will be able to analyze whether backchannels indeed depend on the extraversion personality trait. 

After watching all the videos, the participants were presented with a questionnaire meant to test the three hypotheses mentioned above. In particular, we asked the participants to 
\begin{itemize}
    \item Rank the three policies in order based on the quality of backchannels, taking into consideration factors like the placement, frequency, and selection of backchannel signals.
    \item Rate each one on a Likert scale based on the above parameters.
    \item Identify the personality traits of the avatars based on the videos.
\end{itemize}

Post survey analysis of the participants' responses lead to some exciting findings which supported our hypothesis:

\begin{figure}
    \centering
    \includegraphics[width=\columnwidth]{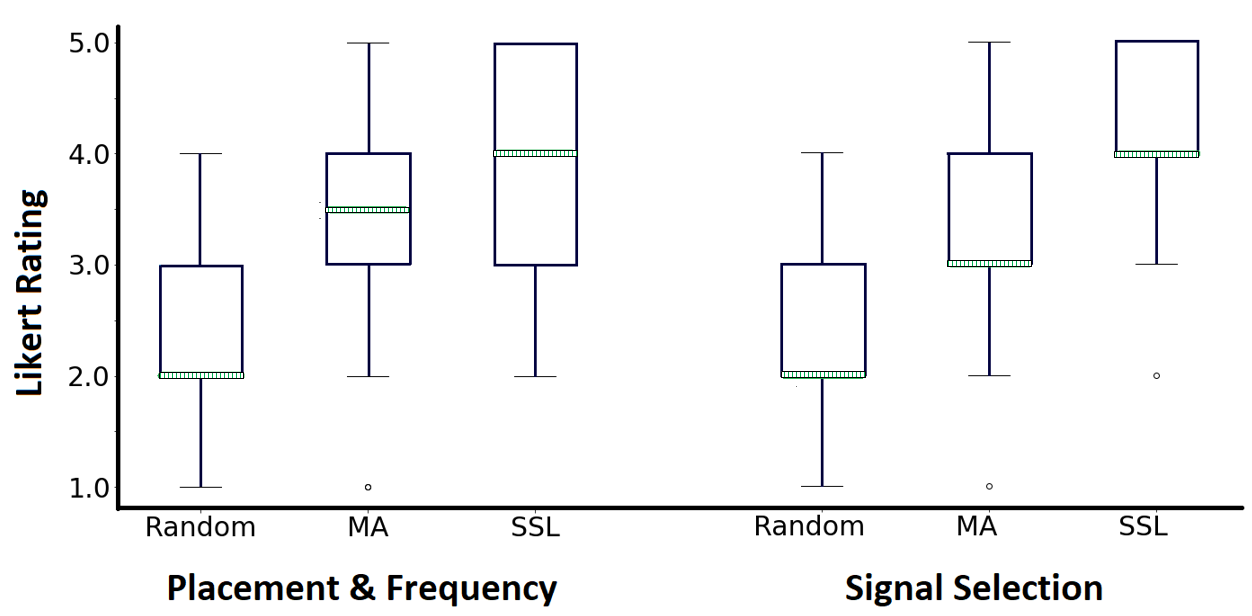}
    \caption{Box \& Whisker plot for the Likert scale ratings (where $1$ is \textit{very poor} and $5$ is \textit{very good}) given by the participants to each of the three policies, based on $(i)$ placement \& frequency, and $(ii)$ signal type selection. }
       \Description{Box & Whisker plot for the Likert scale ratings (where 1 is very poor and 5 is very good) given by the participants to each of the three policies, based on (i) placement & frequency (P&F), and (ii) signal type selection (SS). The boxes for MA-policy and SSL-policy do not overlap with the Random-policy (for both P&F and SS). This indicates that the former two certainly received better ratings than the later. The boxes for MA and SSL overlap in P&F, with the median line for SSL slightly above MA’s. This indicates that there is likely to be a difference between the two sets of ratings, even though it may not be significant (as found from the test). For Signal Selection, the boxes do not overlap at all, indicating a difference. 
        }
    \label{boxplot_qa}
\end{figure}

\begin{enumerate}

    \item \textbf{Models from both the manually-annotated and semi-supervised settings performed better than random:} $95\%$ of the participants ranked our prediction models to be better than the random prediction policy. In addition, the Likert scale ratings suggested that our models performed \textit{significantly} better than the random policy, both in terms of the placement \& frequency, as well as the signal selection. We used the Wilcoxon signed-rank test, and the corresponding p-values are [\textbf{\texttt{P}}- placement \& frequency, \textbf{\texttt{S}}- signal selection, $*$ indicates significant difference at 95\% confidence interval]:  
    
    \begin{itemize}
        \item \textbf{\texttt{MA}} $>$ \textbf{\texttt{Random}}: 0.0019 (\textbf{\texttt{P}}*); 0.0013 (\textbf{\texttt{S}}*) 
        \item \textbf{\texttt{SSL}} $>$ \textbf{\texttt{Random}}: 0.00011 (\textbf{\texttt{P}}*); 0.0001 (\textbf{\texttt{S}}*) 
    \end{itemize}
    
    A box-plot for these ratings is shown in Figure~\ref{boxplot_qa}. Notice how the boxes for \textbf{\texttt{MA}} and \textbf{\texttt{SSL}} do not overlap with \textbf{\texttt{Random}}. This also indicates that the former two certainly received better ratings than the later. This validates our first hypothesis \textbf{\texttt{[H1]}} that with a data-driven approach, we were able to emit more natural and human-like backchannels. 
    
    \item \textbf{Prediction models trained on semi-supervised labels produced more natural backchannels:} Comparing the ratings provided for the \textbf{\texttt{MA}} and the \textbf{\texttt{SSL}} policies, we found that in terms of the frequency and placement (\textbf{\texttt{P}}), there was no \textit{significant} difference between the two (p-value=$0.1024$). For signal selection (\textbf{\texttt{S}}), the ratings suggested that \textbf{\texttt{SSL}} was significantly better than \textbf{\texttt{MA}} (p-value=$0.0122$). The box-plot in Figure~\ref{boxplot_qa} also aligns with these results. In case of Placement \& Frequency, the boxes for \textbf{\texttt{MA}} and \textbf{\texttt{SSL}} overlap, with the median line for \textbf{\texttt{SSL}} slightly above \textbf{\texttt{MA}}'s. This indicates that there is likely to be a difference between the two sets of ratings, even though it may not be significant (as found from the test). For Signal Selection, the boxes do not overlap at all, indicating a difference.
    
    Furthermore, as a part of the questionnaire, we also asked the participants to compare the response videos generated by these two prediction models. Of the total, $60\%$ of the participants found the backchannel responses emitted by the \textbf{\texttt{SSL}} model more natural than the \textbf{\texttt{MA}} model (\textbf{\texttt{SSL}} $>$ \textbf{\texttt{MA}}), and $15\%$ observed no perceptible difference between the two (\textbf{\texttt{SSL}} $\sim$ \textbf{\texttt{MA}}).     This also aligns with our quantitative prediction results where the proposed \textbf{\texttt{SSL}} model was able to reach $\sim 95\%$ of the latter's performance. Thus, the two models were, both qualitatively and quantitatively, very similar.
    
    This confirms our second hypothesis \textbf{\texttt{[H2]}} with most of the participants finding the proposed model (\textbf{\texttt{SSL}}) similar (or more natural) to the \textbf{\texttt{MA}} model. 
    
    \item \textbf{Karan and Arjun's extraversion traits were perceptible:} $80\%$ of the participants were able to accurately identify which of the two virtual-human characters was introvert and extrovert. This confirms our third and final hypothesis \textbf{\texttt{[H3]}} that the type of backchannel signals emitted by an individual indeed depends on their personality. 
\end{enumerate}

\section{Discussion}
\noindent \textbf{Relation to prior work and some interesting findings:} Our quantitative and qualitative evaluations in the previous section greatly validate how semi-supervision can be extremely useful in designing human-like ECAs, by focusing on the task of listener backchannel prediction. We found that with just 25\% of manually annotated data (~175 minutes), we were able to train a backchannel prediction system that performed comparably well, and even better in terms of some parameters, as the one trained using 100\% data. Most of the prior works, even the most recent ones, including \cite{goswami2020social, park2017, Hara2018PredictionOT, Ruede2019}, have depended on a large amount of annotation. We believe that this observation can significantly benefit the HCI community. Furthermore, studies have shown that backchannel responses vary greatly with culture \cite{10.2307/4168001, HEINZ20031113}, and most of the prior studies have focussed primarily on the American and European population in this regard. In this work, we worked with subjects of Indian origin, and therefore, our work holds cultural significance as well. 

A particularly interesting finding from Section~\ref{user_study_sec} was that most participants found the backchannel responses generated by the \textbf{\texttt{SSL}} policy more `natural' than the \textbf{\texttt{MA}} policy. We hypothesise this could be attributed to some form of label noise: After annotations, we only took those positive instances where at least two raters agreed, and the rest were discarded. Our negative samples overlapped with those instances where only 1 of the raters had annotated a positive BC. The \textbf{\texttt{SSL}} model could be learning to predict these backchannels. In fact, in our demo videos, we found two such instances. Starting with a small amount of seed data, the SSL model could have been learning these instances as well (given there are some hints of BC), which could explain the observation that even though SSL performed comparably with \textbf{\texttt{MA}} quantitatively, but seemed more natural to the participants in the subjective study. \vspace{2mm}

\noindent \textbf{Limitations}: We would also like to highlight some limitations of our work, which can form the basis for future studies. First, we only used the participants' visual and acoustic features and did not include the content of the conversation itself for predicting backchannels. The main reason for this was that the primary language used in the \textit{Vyaktitv} dataset was Hindi. The English translations for the dialogues were not available, making the use of state-of-the-art NLP techniques non-trivial. In another aspect, our qualitative user case study involved a short 4-minute long compilation of speakers from the \textit{Vyaktitv} dataset. A more rigorous analysis could follow by deploying the models and the avatars as a real time-system. Finally, we did not annotate eye blinks/gaze \cite{homke2018eye} as backchannel signals, primarily because they were not as apparent in the dataset. This could be a cultural difference as well.  \vspace{2mm}

\noindent\textbf{Ethical Consideration}: When developing data-driven systems leveraging data from human-subjects, it becomes imperative that we respect their privacy boundaries. We want to state that no Personally Identifiable Information (PII)  was used while training or evaluating the system. We also complied with the agreements in the \textit{Vyaktitv} dataset to ensure the safe use of data.

\section{Conclusion}
\label{conclusion}

In this work, we confirmed the feasibility of using semi-supervised learning to (semi-) automate the process of identifying and labeling listener backchannel instances (both opportunities and the associated signals) from conversations. We used a Hindi peer-to-peer conversation-based multimodal dataset \textit{Vyaktitv}, for our experiments. However, the methodology proposed in the study is general and can be adapted for other conversational datasets as well. Quantitative evaluation alongside a subjective analysis in the form of a user study strongly validated our hypothesis that- prediction models trained using semi-supervised labels performed comparably with those using manually annotated labels. Furthermore, we statistically and qualitatively confirmed that the type of backchannel signals emitted are intimately linked to an individual's personality (extraversion in particular).     

Future work directions include validating the scope of semi-supervised learning for listener backchannel prediction on other datasets. Other parallel tasks, like listener disengagement prediction, can also be similarly performed. Methodologically, future studies could explore by devising heuristics and using weak supervision based techniques to identify backchannels. Our observations from analyzing the impact of personality on the type (modality) of backchannel response also open some new research questions; for instance- can we also analyze if the frequency of backchannels emitted by different individuals depend on personality \cite{bevacqua2012listener}? Furthermore, can we use these findings to embed personality into a virtual human further? These are some exciting lines future researchers can look into by conducting more extensive analysis.  

\section*{Acknowledgements} 
Jainendra Shukla and Rajiv Ratn Shah are partly supported by the Infosys Center for AI and Center for Design and New Media at IIIT Delhi. In addition, Rajiv Ratn Shah is also partly supported by the ECRA Grant (ECR/2018/002776) by SERB, Government of India. Finally, we would like to thank the annotators, and all the participants who took part in our case study for their time and efforts.

\bibliographystyle{ACM-Reference-Format}
\bibliography{acmart}

\end{document}